\documentclass[amsmath,amssymb,aps,prl,superscriptaddress,twocolumn]{revtex4-1}
\usepackage{graphicx}
\usepackage{dcolumn}
\usepackage{bm}
\usepackage[usenames]{color}
\usepackage{comment}
\usepackage{subfigure}

\usepackage{hyperref}
\usepackage{bookmark}
\usepackage{mathtools}
\begin{document}

\title{Topological orders competing for the Dirac surface state in FeSeTe surfaces
}

\author{Xianxin Wu}
\affiliation{Department of Physics, the Pennsylvania State University, University Park, PA, 16802}
\affiliation{Beijing National Laboratory for Condensed Matter Physics, and Institute of Physics, Chinese Academy of Sciences, Beijing 100190, China}
\author{Suk Bum Chung}
\affiliation{Department of Physics, University of Seoul, Seoul 02504, Korea}
\affiliation{Natural Science Research Institute, University of Seoul, Seoul 02504, Korea}
\affiliation{School of Physics, Korea Institute for Advanced Study, Seoul 02455, Korea}
\author{Chaoxing Liu}
\affiliation{Department of Physics, the Pennsylvania State University, University Park, PA, 16802}
\author{Eun-Ah Kim}
\email{eun-ah.kim@cornell.edu}
\affiliation{Department of Physics, Cornell University, Ithaca, New York 14853, USA}

\begin{abstract}
FeSeTe has recently emerged as a leading candidate material for the two-dimensional topological superconductivity (TSC).
Two reasons for the excitement are the high $T_c$ of the system and the fact that the
Majorana zero modes (MZMs) inside the vortex cores
live on the exposed surface rather than at the interface of a heterostructure as in the proximitized topological insulators.
However, the recent scanning tunneling spectroscopy data have shown that, contrary to the theoretical expectation, the MZM does not exist inside every vortex core. Hence there are ``full'' vortices with MZMs and ``empty'' vortices without MZMs. Moreover the fraction of ``empty'' vortices increase with an increase in the magnetic field.
We propose the possibility of two distinct gapped states competing for the topological surface states in FeSeTe: the TSC and half quantum anomalous Hall (hQAH). The latter is promoted by magnetic field through the alignment of magnetic impurities such as Fe interstitials. When hQAH takes over the topological surface state, the surface will become transparent to scanning tunneling microscopy and the nature of the vortex in such region will appear identical to what is expected of the vortices in the bulk, i.e., empty. Unmistakable signature of the proposed mechanism for empty vortices will be the existance of chiral Majorana modes(CMM) at the domain wall between a hQAH region and a TSC region. Such CMM should be observable by observing local density of states along a line connecting an empty vortex to a nearby full vortex.

\end{abstract}

\maketitle

 {\it Introduction --}
One particularly exciting feature of the topological insulator (TI)
its potential to host
the Majorana zero mode (MZM), which
has led to many proposals \cite{Fu2007B, Qi2008B, Fu2008L, Fu2009L} and attempts \cite{MXWang2012, Veldhorst2012, Kurter2015, JPXu2015, HHSun2016} to realize MZM through introducing superconducting gap to the TI surface state.
Early works focused on
introducing topological superconductivity (TSC) through proximity effect \cite{Fu2008L, Akhmerov2009, Hosur2011, KMLee2014}.  More recently, the prospect of FeSeTe
possessing at its surface the equivalent of TI surface state
with superconducting gap proximity induced by the high T$_c$ intrinsic bulk superconductivity
raised much enthusiasm \cite{NNHao2014, ZJWang2015, XXWu2016, GXu2016}. More recently it has been recognized that such state
possesses a higher order topology \cite{RXZhang2019,Gray2019,wu2019high,ZhangRX-PRL20192}.

Intensive experimental investigations of FeSeTe confirmed the existence of Dirac surface state  in the normal state above T$_c$ \cite{PZhang2018}. The predicted evidence 
for the MZM in the vortex core of superconducting state was
the zero-bias peak  in scanning tunneling microscopy (STM). Indeed, the STM is a particularly suitable probe for the MZM in this material as it would exist at the surface \cite{Hosur2011, KMLee2014}. Despite several observations of a zero-bias peak in cores of some vortices \cite{JXYin2015, DFWang2018, LYKong2019, SYZhu2020}, an apparent contradiction to the prediction has also been observed in the increasing fraction of ``empty'' vortices without a zero-bias peak upon the increase in magnetic field \cite{XChen2019,Machida2019}.
A careful study \cite{XChen2019} revealed that the ``empty'' vortices cannot be accounted for by a simple picture of pair-wise annihilation of MZM between two near-by vortices.  Although Ref.~\cite{chiu2020scalable} showed that a model allowing for long-range interaction among MZM's far separated can in principle explain the ``empty'' vortices, an alternative explanation with simpler starting point and a falsifiable prediction is desirable.

\begin{figure}
    \centering
    \includegraphics[width=.48\textwidth]{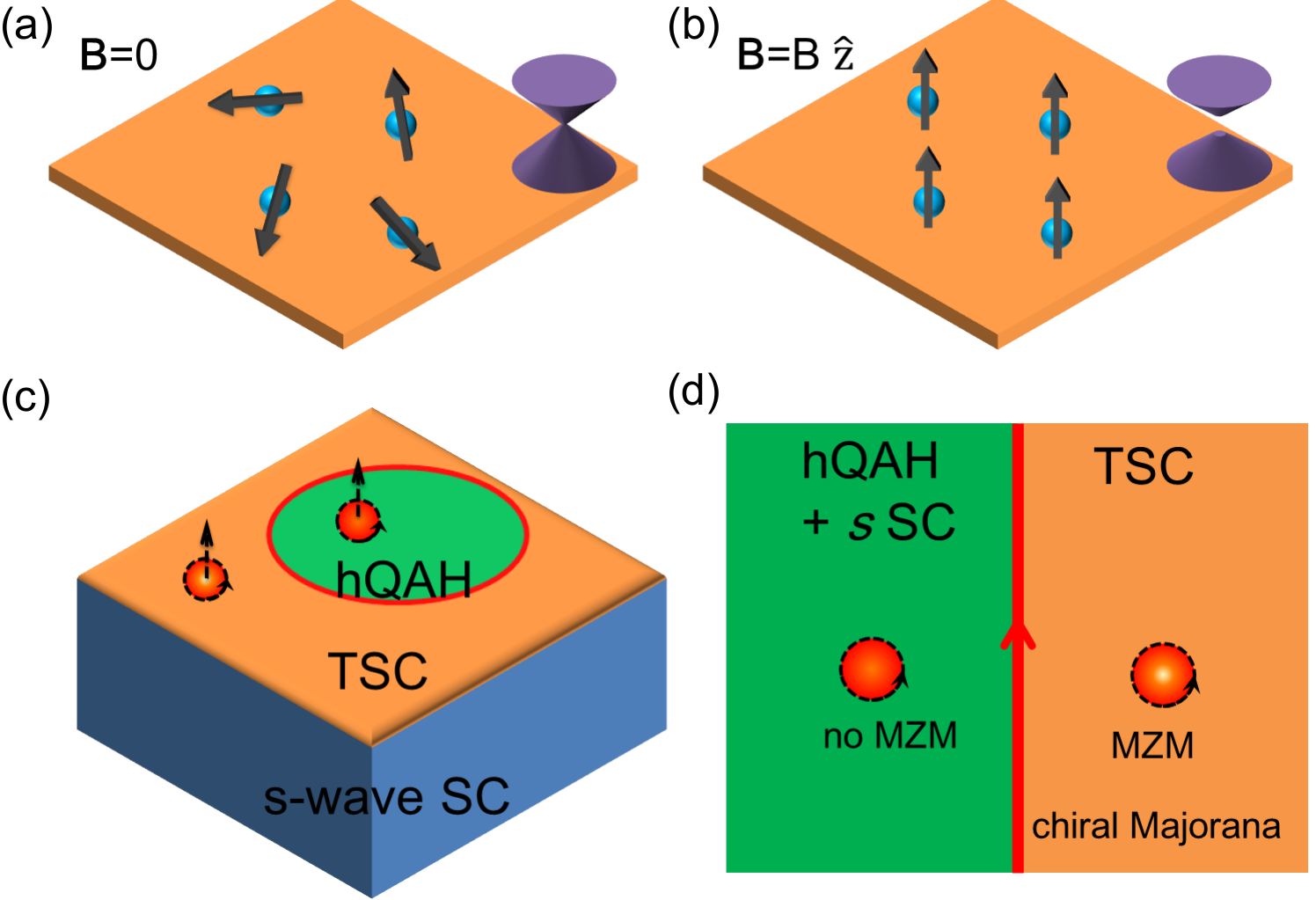}
    \caption{(a) Gapless Dirac surface state with random magnetic moments of Fe interstitials and (b) the gaped surface state when magnetic moments are aligned by external magnetic fields. (c) Domain wall between the TSC region (with dominant superconducting gap) and the hQAH region (with dominant magnetic gap) on the surface of FeTeSe. (d) MZM exists at the vortex core (the red spots) in the TSC region, but not in the hQAH region. The CMM exists at the boundary between the TSC and hQAH regions.   }
    \label{fig:idea}
\end{figure}

 Here we provide an alternative interpretation of the observed "empty" vortices based on the role of the magnetic field on aligning local moments of Fe-interstitials. Our main physical picture is summarized in Fig.~\ref{fig:idea}a-d. As it is known from the study of magnetic dopants added to TI surface states, the exchange field from magnetic impurities also gap the TI surface state to form the half quantum anomalous Hall (hQAH) state with the half-integer quantization of Hall conductivity \cite{Qi2008B, CXLiu2008L, CZChang2013, PWei2013}  {\color{blue}(Fig.~\ref{fig:idea}a -\ref{fig:idea}b)}
Uneven distribution of interstitials can nucleate the hQAH regions on the surface of FeSeTe when the moments get aligned with magnetic fields {\color{blue}(Fig.~\ref{fig:idea}c)}, preventing TSC to form in that very region. Such hQAH surface state will reveal the bulk superconductivity to STM and the vortices penetrating hQAH surface will show properties of the bulk superconducting state with topologically trivial the $s\pm$ pairing \cite{Sprau2017, DFLiu2018}, i.e., becoming ``empty''. With increasing magnetic fields, more hQAH regions are nucleated on the surface of FeSeTe, thus providing a natural explanation of the increasing faction of empty vortices observed in experiments. Interestingly, it has been known that a boundary between hQAH and TSC should host a chiral Majorana mode (CMM) \cite{Fu2009L, Qi2010B, Chung2011, JWang2015}.
Hence our key prediction is that the MZM that would have been in the vortex core transforms into the CMM located at the boundary between the hQAH and TSC on the surface of FeTeSe{\color{blue}(Fig.~\ref{fig:idea}d)}. In the rest of this Letter, we first present our proposal using
a low energy effective theory and then support it
with a  numerical simulation on a microscopic  model.

 {\it Exchange field and low energy effective theory --}
Consider the low energy effective theory for the topological Dirac surface state in FeSeTe.
As noted by Jiang {\it et al.} \cite{KJiang2019}, the interstitial Fe atoms can provide magnetic impurities in Fe(Te$_{0.55}$Se$_{0.45}$). Although the impurity moments will point in random direction at zero-field (Fig 1a), the external field applied to create vortices would align the impurity moments (Fig 1b). In the regions with higher concentration of aligned impurity moments, the exchange field generated by these moments would couple to the topological surface state as in magnetically doped TI \cite{QLiu2009, Henk2012, Rosenberg2012, Efimkin2014, YLChen2010, Wray2011}.
Such exchange coupling can be captured by
$H_{ex}({\bf r}) =  -\mathcal{J}_0 \sum_i {\bf S}_i \cdot {\bf s} \delta({\bf r}-{\bf r}_i)$,
 where ${\bf s}=\frac{\hbar}{2}{\bf \sigma}$ is the surface state electron spin, ${\bf S}_i$ and ${\bf r}_i$ are the spin and location, respectively, of the Fe interstitial and $\mathcal{J}_0$ is the coupling constant. This exchange field will be heterogeneous depending on the distribution of the interstitials. We consider the mean field approximation for the exchange field, leading to the form
 $H_{ex}({\bf r}) =  -{\bf I_{ex}}({\bf r})\cdot{\bf \sigma}$, where ${\bf I_{ex}}({\bf r})=\frac{\mathcal{J}_0\hbar}{2} \sum_i \langle {\bf S}_i \delta({\bf r}-{\bf r}_i)\rangle_{local}$ is a smoothly varying field with $\langle ...\rangle_{local}$ representing the average over a small region for Fe moments.
 In an ordinary topological insulator,
 such heterogeneous exchange field should result in hQAH effect with spatially varying gaps for the Dirac surface state~\cite{IHLee2015}. However,
 non-topological bands crossing the Fermi Surface will mask hQAH states in the normal state of Fe(Te$_{0.55}$Se$_{0.45}$).

Once the system develops superconductivity, the hQAH and TSC can compete as
the two possible ways of gapping the Dirac surface state.
Moreover, the hQAH region will reveal itself by leaving the bulk superconductivity bare when the exchange gap dominates over the superconducting gap. This can be captured by
the BdG Hamiltonian for the Dirac surface state with both exchange field and the $s$-wave pairing in the basis $(c_{{\bf k},\uparrow}, c_{{\bf k},\downarrow}, c^\dagger_{-{\bf k},\downarrow},-c^\dagger_{-{\bf k},\uparrow})^T$:
\begin{equation}
\mathcal{H}_{BdG} = (v{\bf k} \cdot {\bm \sigma} - \mu)\tau_z - I_{ex} \sigma_z + \Delta\tau_x,
\label{EQ:BdGHam}
\end{equation}
where $\sigma_i$ and $\tau_i$ are the Pauli matrices in the spin space and particle-hole space, respectively. Here we assumed an $s$-wave gap to be real and only consider exchange field along the z direction.
It is straightforward to find upon increase in the exchange term, the superconducting gap for BdG quasiparticles closes at the critical exchange field strength of \cite{Sato2009, Sau2010L, Alicea2010}
\begin{equation}
    I^2_{ex,c}=|\Delta|^2+\mu^2.
\label{EQ:critical}
\end{equation}

 When $|I_{ex}|< |I_{ex,c}|$, the TSC dominates to support the vortex core MZM, which can be explicitly obtained by choosing
$\Delta\tau_x \to |\Delta|(\tau_x \cos \theta - \tau_y \sin \theta)$ substitution ($\theta$ is the azimuthal angle), which places a superconducting vortex at the origin. The zero mode we obtain for $|I_{ex}|<|\mu|$ \cite{Fu2008L},
\begin{equation}
\left[\begin{matrix} \psi_\uparrow ({\bf r}) \\ \psi_\downarrow ( {\bf r}) \\ \psi^\dagger_\downarrow ({\bf r}) \\ -\psi^\dagger_\uparrow({\bf r})\end{matrix}\right]\!=\!\frac{e^{-\int^r_0 dr' \frac{|\Delta|}{\hbar v}}}{(\mu^2\!-\!I_{ex}^2)^{\frac{1}{4}}}
\left[\begin{matrix} e^{-i\frac{\pi}{4}}\sqrt{\mu\!+\!I_{ex}} J_0 \left(\frac{\sqrt{\mu^2\!-\!I_{ex}^2}}{\hbar v}r\right) \\ e^{i\frac{\pi}{4}}e^{i\theta}\sqrt{\mu\!-\!I_{ex}} J_1 \left(\frac{\sqrt{\mu^2\!-\!I_{ex}^2}}{\hbar v}r\right) \\ e^{-i\frac{\pi}{4}}e^{-i\theta}\sqrt{\mu\!-\!I_{ex}} J_1 \left(\frac{\sqrt{\mu^2\!-\!I_{ex}^2}}{\hbar v}r\right) \\ -e^{i\frac{\pi}{4}}\sqrt{\mu\!+\!I_{ex}} J_0 \left(\frac{\sqrt{\mu^2\!-\!I_{ex}^2}}{\hbar v}r\right)\end{matrix}\right]
\end{equation}
where $J_l$ is the $l$-th Bessel function of the first type, reduces the Fu-Kane vortex zero mode by setting first $I_{ex}=0$ and then $\mu=0$ \cite{Fu2008L}. It can also be generalized to $|I_{ex}|>|\mu|$ using $J_l (ix) = i^n I_l(x)$ for $x \in \mathbb{R}$, where $I_l$ is the $l$-th modified Bessel function of the first type, provided, however, that $|\Delta(r \to \infty)|>\sqrt{I_{ex}^2-\mu^2}$, {\it i.e.}
$|I_{ex}|< |I_{ex,c}|$, as can be seen from the asymptotic forms for the large real arguments, 
$I_l (x) \sim e^x/\sqrt{2\pi x}$. 

 On the other hand, when $|I_{ex}|> |I_{ex,c}|$, hQAH dominates without the vortex core MZM. The domain wall CMM can be demonstrated by setting 
$\mu=0$ with the domain wall at $y=0$ arising from $I_{ex}(y) = I_0 \Theta(y)$ and $\Delta=\Delta_0 \Theta(-y)$ will be considered, {\it i.e.}
\begin{equation}
\mathcal{H}_{BdG} = v{\bf k} \cdot {\bm \sigma}\tau_z - I_0 \Theta(y) \sigma_z + \Delta_0 \Theta(-y)\tau_x;
\label{EQ:domain}
\end{equation}
for $I_0=\Delta_0>0$, it is straightforward to show the existence of the domain wall CMM
\begin{equation}
\left[\begin{matrix} \psi_\uparrow ({\bf r}) \\ \psi_\downarrow ( {\bf r}) \\ \psi^\dagger_\downarrow ({\bf r}) \\ -\psi^\dagger_\uparrow({\bf r})\end{matrix}\right]=\sqrt{\frac{\Delta_0}{vL_x}}e^{ik_x x} e^{-\Delta_0 |y|/v}\left[\begin{matrix} 1/2 \\ 1/2 \\ 1/2 \\ -1/2 \end{matrix}\right]
\end{equation}
with the eigenenergy $E_{k_x}=vk_x$.

{\it Microscopic model --} Next we will support our results by the numerical simulations on the bulk model of FeSeTe system. For FeSeTe bulk system, the topological phase is attributed to the band inversion between two states with opposite parities at Z point. Taking $|S^{+}, +\frac{1}{2}\rangle$, $|S^{+}, -\frac{1}{2}\rangle$, $|P^{-}, \frac{1}{2}\rangle$ and $|P^{-}, -\frac{1}{2}\rangle$ as the basis at Z point, the topological electronic structure can be described by the Hamiltonian in a 3D lattice $H_{TI}=\sum_{\bm{k}}\psi^\dag_{\bm{k}}\mathcal{H}_{TI}(\bm{k})\psi_{\bm{k}}$ and Hamiltonian matrix reads
\begin{eqnarray}
\mathcal{H}_{TI}(\bm{k})=\eta_x\bm{d}\cdot\bm{\sigma}+M_{\bm{k}}\eta_z-\mu,
\end{eqnarray}
where $\psi^\dag_{\bm{k}}=(c^\dag_{S\bm{k}\uparrow},c^\dag_{S\bm{k}\downarrow},c^\dag_{P\bm{k}\uparrow},c^\dag_{P\bm{k}\downarrow})$, $M_{\mathbf{k}}=M_0+m_{z}\cos k_z+m_{x}(\cos k_x+\cos k_y)$ and $d_i=2t_i\sin k_i$ ($i=x,y,z$). Here $\bm{\eta}$ are Pauli matrices in the orbital space. The mass term at $\Gamma$ and $Z$ points are $M_0+m_z+2m_x$ and $M_0-m_z+2m_x$. Let us take $m_0=-m_z=m_x$, the above model describes a strong topological insulator phase with a band inversion at $Z$ point if $-3<\frac{M_0}{m_0}<-1$ is satisfied.

We extend the Hamiltonian to include superconductivity and exchange field from impurities, the BdG Hamiltonian is $H_{BdG}=\frac{1}{2}\sum_{\bm{k}}\Psi^\dag_{\bm{k}}\mathcal{H}^{TI}_{BdG}(\bm{k})\Psi_{\bm{k}}$ with $\Psi_{\bm{k}}=[\psi^\dag_{\bm{k}},\psi^T_{-\bm{k}}(-i\sigma_y)]$ and the Hamiltonian matrix reads,
\begin{equation}
\mathcal{H}^{TI}_{BdG}(\mathbf{k})  = \left(\begin{array}{cc}
\mathcal{H}_{TI}(\bm{k})-I_{ex}\sigma_z & \Delta_s  \\
\Delta^\dag_s & -\mathcal{H}_{TI}(\bm{k})-I_{ex}\sigma_z \\
\end{array}\right),
\label{eqBdG}
\end{equation}
where $\Delta_s$ is the intra-orbital spin singlet pairing. In the absence of exchange field, the (001) surface states will be gapped by superconductivity and form an effective $p+ip$ pairing, where Majorana modes can be trapped in a vortex core of the surface (as described by Eq.~\ref{EQ:BdGHam} with $I_{ex}=0$). We then study the effect of exchange field on the (001) surface states by adopting the above Hamiltonian with open boundary condition along $z$ direction.

 The microscopic model reproduces the topological phase transition of the low energy effective theory. Fig.~2 demonstrates the existence of the topological phase transition of the surface states by fixing the pairing potential and increasing the exchange field strength. 
In Fig.~2a, with zero exchange field, the surface state is gapped by superconducting pairing. When the exchange field strength reaches the critical strength which is equal to the superconducting gap for $\mu=0$, the gap of the surface states closes (Fig.~2b), consistent with the condition of Eq.~\ref{EQ:critical}. With further increasing exchange field, the surface state gap reopens and the system is driven into the hQAH state (Fig.~2c).

\begin{figure}
    \centering
    \includegraphics[width=.48\textwidth]{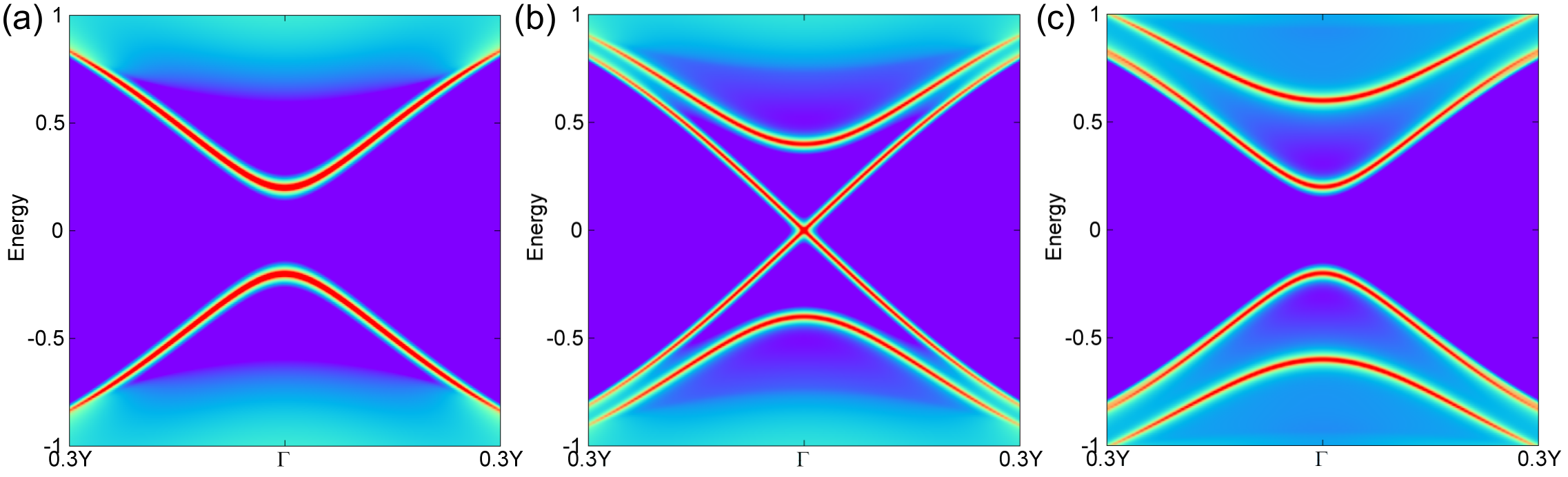}
    \caption{Topological phase transition for (001) surface states with increasing exchange field $I_{ex}$: (a) $I_{ex}=0$, (b) $I_{ex}=-0.2$ and (c) $B_z=-0.4$. The blue and red dots represent surface state at top and bottom surfaces, respectively. The adopted parameters are: $t_{x,y}=t_z=0.5$, $M_0=2.5$, $m_x=-m_z=-1.0$, $\Delta_0=0.2$ and $\mu=0$.}
    \label{fig:my_label}
\end{figure}

 Next we turn to how exchange field affects the vortex core MZM in topological surface state superconductivity. We introduce a vortex located at the center of the system by setting $\Delta_s(r)=|\Delta_s(r)|e^{i\theta}$ and adopt the Hamiltonian with open boundary conditions along the $x,y,z$ directions. A lattice size of $17\times17\times16$ is chosen for the following numerical calculations. The exchange field is only restricted to the top (001) surface of the system.

 With the above sample configuration, Figs. 3a and b show the distribution of the zero-energy local density of states on the bottom and top surfaces, respectively, for the exchange field exceeding the critical strength defined by Eq. \ref{EQ:critical}. One can see that an ``empty vortex'' appears on the top surface in Fig. 3b, in sharp contrast to the ``full vortex'' on the bottom surface where there is no exchange field in Fig 3a. At the core of a full vortex, there is a well-defined MZM with zero-bias peak in the local density of states. On the other hand, the MZM is absent at the core of an empty vortex. Despite of the absence of MZM in the vortex core, the edges of the top surface under exchange field show a large amount of density of states that depict the presence of edge CMM. In Supplementary Materials, we study the profile evolution of zero-energy local density of states on the top surface with increasing magnetic fields, from which one find that the localized MZM gradually extends outside of the vortex and becomes localized on the the edges of (001) surface.

 {\it Experimental prediction --}
Based on our results that have been well established by both the effective theory and microscopic bulk model, a natural prediction is the existence of the domain wall CMM between an empty vortex and a full vortex. Consider an experimental setup shown in Fig. 1c, in which two vortices are located at the TSC and hQAH regions, respectively. Fig.4a displays the spatial profiles of zero-energy states in the vicinity of a full vortex (left) and an empty vortex (right) and Fig.4b shows the progrssion of the local density of states (LDOS) as a tip marches from a full vortex to an empty vortex.  
Experimentally, one can implement an STM measurement of LDOS along the line connecting a full vortex (indicating the TSC state) and an empty vortex (indicating the hQAH state). As shown in Fig. 4b, a zero-bias peak is expected to exist in the intermediate region without any vortex and can be attributed to the existence of the CMM at the domain wall between the hQAH and TSC regions. This chiral Majorana mode always possessing a zero-energy state is distinct from a normal chiral mode  and the energy spectrum is related to the circumference of the region with Zeeman field (see SM). With a large  thermal smearing in STM measurements, the LDOS at the domain wall exhibits a broad peak around zero energy.  While the external magnetic field cannot gap out the CMM, changing its magnitude will shift the location of the CMM as the hQAH region expands while the TSC region contracts or vice versa.

\begin{figure}
    \centering
    \includegraphics[width=0.45\textwidth]{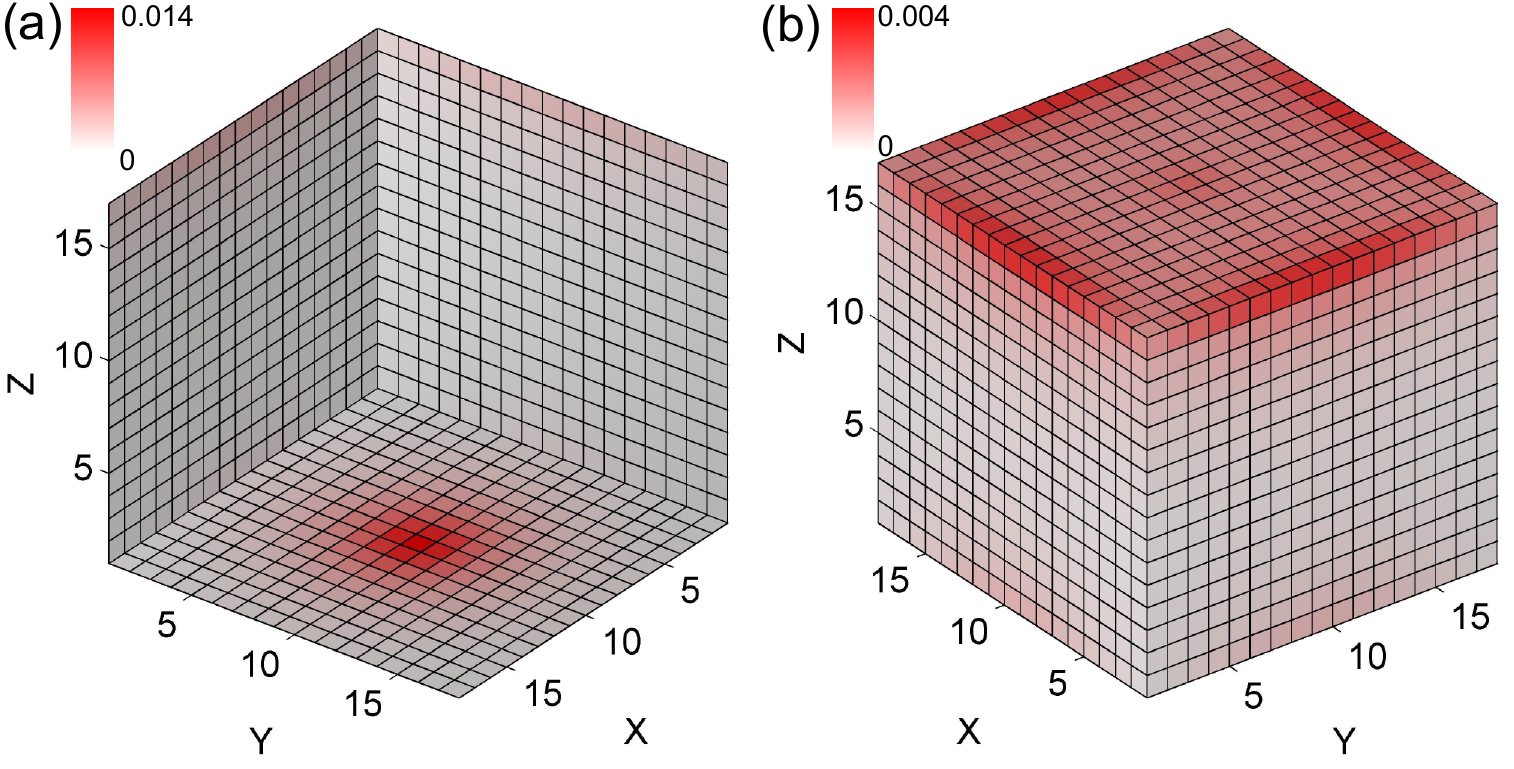}
    \caption{The 3D profiles of MZMs for (001) surface on a $17\times17\times16$ lattice with $I_{ex}=-0.4$, $\Delta_0=0.2$ and $\mu=0$. There is a localized Majorana in the vortex and chiral Majorana modes localized on edges on bottom and top surface, respectively.}
    \label{fig:my_label}
\end{figure}

\begin{figure}
    \centering
    \includegraphics[width=.45\textwidth]{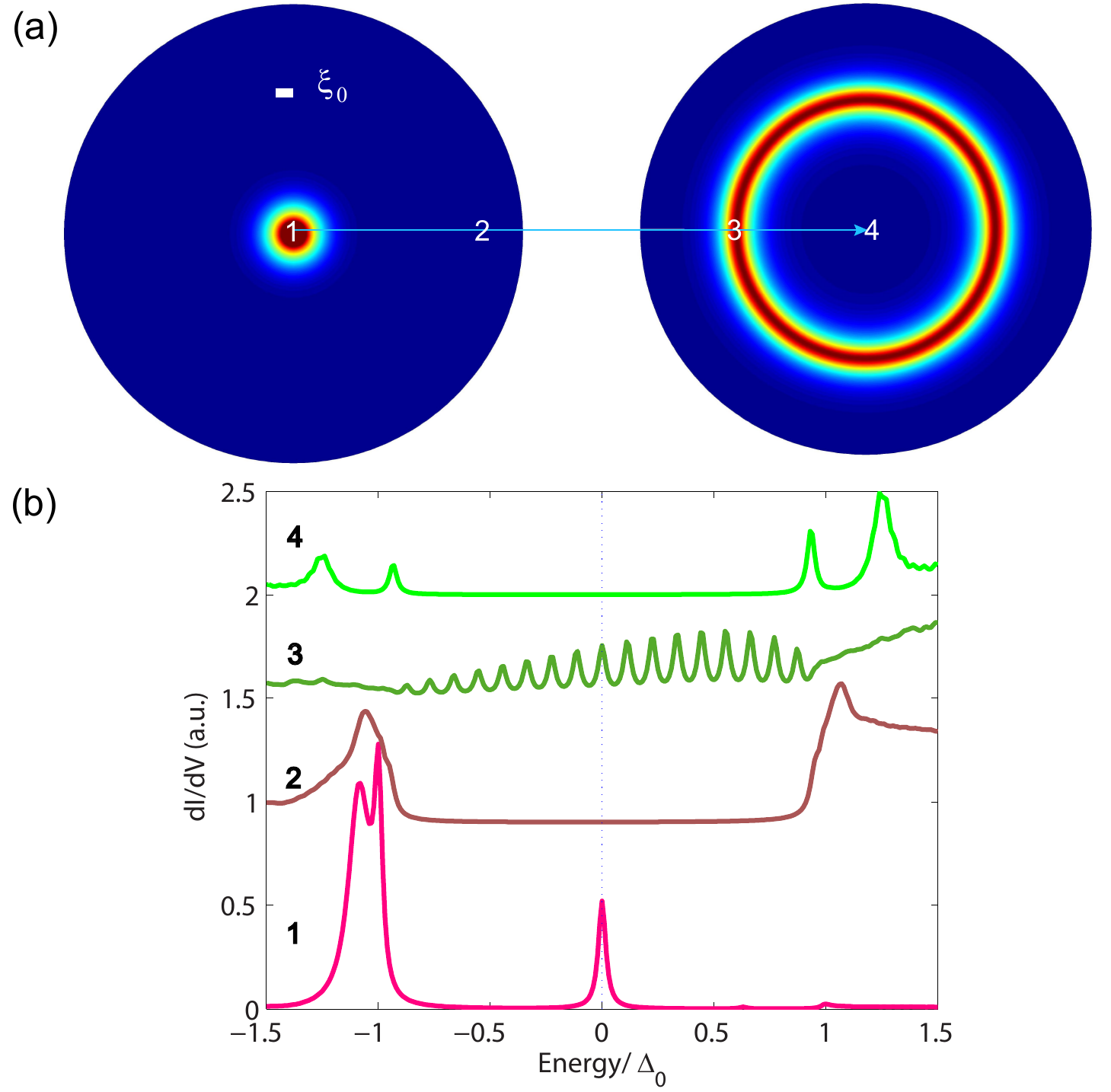}
    \caption{The profiles of MZMs at top and bottom (001) surfaces (a) and position-dependent local density of states between a normal and an ``empty" vortices (b). }
    \label{fig:my_label}
\end{figure}

 {\it Conclusion--}
To summarize, we proposed a new mechanism by which magnetic field can increase the fraction of ``empty'' vortices without MZM in Fe(Te$_{0.55}$Se$_{0.45}$).
Our mechanism is purely {\it local}, {\it i.e.} a vortex is ``empty" because of its intersecting the surface inside the hQAH domain rather than the long-range MZM interaction effects. We postulate that these hQAH domains arise from the alignment of local moments associated with Fe interstitial which produces heterogeneous exchange fields 
exceeding the superconducting gap in isolated puddles. It has been known that there should be the CMM localized at the domain wall between regions with dominant superconducting gap and regions with dominant exchange gap. Through an explicit calculation on a minimalistic lattice model of topological bands, we showed that MZM in the vortex core of topological superconductor transforms into the domain wall CMM upon increase in the exchange field on the region supporting the vortex.

Our proposal is distinct from an earlier proposal in Ref.~\cite{chiu2020scalable} that relies on pair-wise extinction of MZM's through tunneling between vortices. In our proposal, the MZM relocates and extends to the domain wall CMM instead of  disappearing. A clear signature of the proposed mechanism will be the existence of the domain wall CMM between an ``empty'' vortex and a ``full'' vortex which can be detected through STM measurements along a line connecting an ``empty'' vortex to a nearby ``full'' vortex.
Given the clear distinction between the domain wall and the vortex core as shown in Fig.~4b, 
our proposal 
suggests that the CMM detection in Fe(Te$_{0.55}$Se$_{0.45}$) through the STM measurement may be relatively easy compared to the recent transport experiments \cite{QLHe2017, Kasahara2018}.
Another prediction that should be easy to check is that we  anticipate the ``full'' vortices and ``empty'' vortices to segregate as their segregation will represent the regions dominated by TS or by hQAH.

\noindent{\bf Acknowledgements--} We thank Hai-Hu Wen, Tetsuo Hanaguri, and Vidya Madhavan for useful discussions.  EAK was supported by  National Science Foundation (Platform for the Accelerated Realization, Analysis, and Discovery of Interface Materials (PARADIM)) under Cooperative Agreement No. DMR-1539918.
C.X.L acknowledges the support of the Office of Naval Research (Grant No. N00014-18-1-2793), the U.S. Department of Energy (Grant No.~DESC0019064) and Kaufman New Initiative research grant KA2018-98553 of the Pittsburgh Foundation.
SBC acknowledges the support of the National Research Foundation of Korea(NRF) grant funded by the Korea government(MSIT) (No. 2020R1A2C1007554).

\bibliographystyle{apsrev4-1}
\bibliography{FeSeTe}
\appendix
\clearpage
  \begin{widetext}
\section{Evolution of Majorana modes on top (001) surface}

 We include a Zeeman field on the (001) surface to investigate its effect the on Majorana states. With increasing magnetic field, a topological phase transition on (001) surface states will occur, as shown in Fig.3 in the main text. If the magnetic field is large enough (larger than $\sqrt{\Delta_0^2+\mu^2}$), the (001) surface becomes topologically trivial. As the other sides surface states are topologically nontrivial, chiral Majorana modes should occur. Fig.\ref{MZMBz} shows the profiles of Majorana modes as a function of Zeeman field $I_{ex}$. With increasing Zeeman field, the localized Majorana mode at vortex core gradually becomes extended and finally transforms into a chiral Majorana mode (on a $17\times17\times16$ lattice), as demonstrated in Fig.5.

\begin{figure}[tb]
\centerline{\includegraphics[width=1.0\columnwidth]{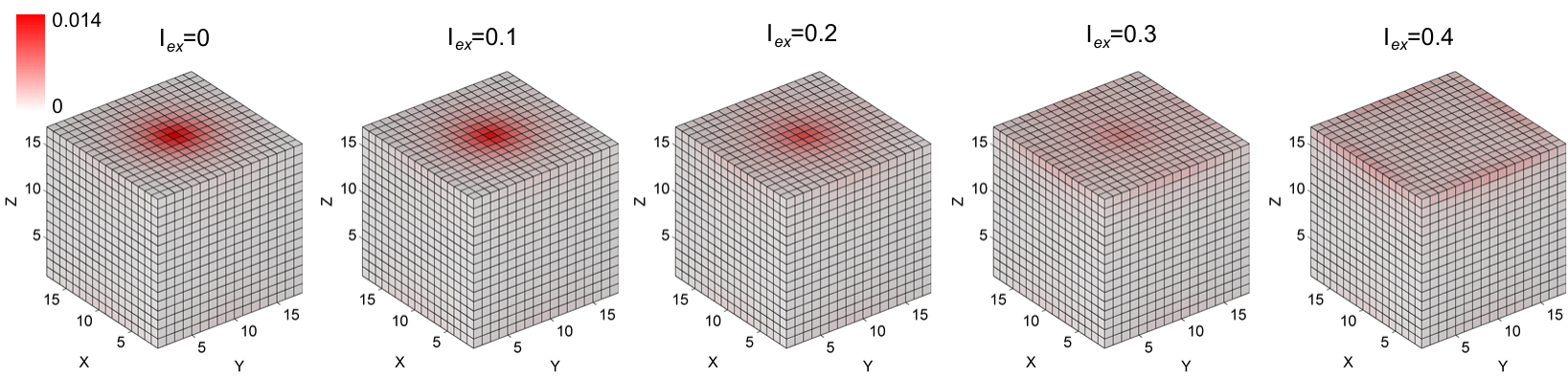}}
\caption{(color online) The profiles of Majorana mode for top (001) surface as a function of $I_{ex}$ (on a $17\times17\times16$ lattice). $\Delta_0=0.2$ and $\mu=0$ are adopted.
 \label{MZMBz} }
\end{figure}

\section{Vortex states in the superconcuting 2D Dirac surface states}
 Now we consider the Dirac surface states on the surface of a topological insulator in proximity to a superconductor with an exchange field $B$. The corresponding BdG Hamiltonian reads,
\begin{eqnarray}
H_s=v_F\tau_z(p_x\sigma_x+p_y\sigma_y)-\mu\tau_z+B\sigma_z+\Delta(r,\theta)\tau_x
&=& \left(\begin{array}{cccc}
-\mu+B & p_x-ip_y & \Delta(r,\theta) & 0\\
p_x+ip_y & -\mu-B & 0  & \Delta(r,\theta) \\
\Delta^\dag(r,\theta) & 0 & \mu+B & -p_x+ip_y \\
0 & \Delta^\dag(r,\theta) & -p_x-ip_y & \mu-B  \\
\end{array}\right),
\end{eqnarray}
where the basis is $\Psi_{\bm{p}}=(c^\dag_{\bm{p}\uparrow},c^\dag_{\bm{p}\downarrow},c_{-\bm{p}\downarrow},-c_{-\bm{p}\uparrow})$. Here $\bm{\tau}$ and $\bm{\sigma}$ are Pauli matrices in Nambu and spin space and the gap function $\Delta(r,\theta)=\Delta_0f(r)e^{in\theta}=\Delta_0 \tanh\frac{r}{\xi_0}e^{in\theta}$ ($n$ is the vorticity of the vortex). With the above basis, the time reversal operation is $\mathcal{T}=-i\sigma_y \mathcal{K}$ and the particle-hole operation $\mathcal{C}=\tau_y \sigma_y\mathcal{K}$. The above Hamiltonian satisfies: $\mathcal{T}H_{s}(\bm{p})\mathcal{T}^{-1}=H_{s}(-\bm{p})$ and $\mathcal{C}H_{s}(\bm{p})\mathcal{C}^{-1}=-H_{s}(-\bm{p})$.  In the real space, we use the substitution $p_{x,y}\rightarrow -i\partial_{x,y}$ and we have the following equations,
\begin{eqnarray}
p_x-ip_y&=&-i(\partial_x-i\partial_y)=-ie^{-i\theta}\partial_r-\frac{e^{-i\theta}}{r}\partial_{\theta},\\
p_x+ip_y&=&-i(\partial_x+i\partial_y)=-ie^{i\theta}\partial_r+\frac{e^{i\theta}}{r}\partial_{\theta},
\end{eqnarray}
As there is a rotational symmetry, the angular momentum is conserved and we can express the above BdG equations as a set of 1D radial equations separated into angular momentum modes. In the following we consider the vortex with $n=1$ and assume the trial wavefunction has the following form,
\begin{eqnarray}
\Psi(r,\theta)=
\frac{e^{i\nu\theta}}{\sqrt{r}}\left(\begin{array}{c}
e^{i\frac{\pi}{4}}u_{\uparrow}(r)\\
e^{i\theta-i\frac{\pi}{4}}u_{\downarrow}(r)\\
e^{-i\theta+i\frac{\pi}{4}}v_{\downarrow}(r)\\
e^{-i\frac{\pi}{4}}v_{\uparrow}(r)\\
\end{array}\right)
=\frac{e^{i\nu\theta-\frac{i\theta}{2}\sigma_z+i\frac{\pi}{4}\sigma_z+\frac{i\theta}{2}\tau_z}}{\sqrt{r}}\left(\begin{array}{c}
 u_{\uparrow}(r)\\
 u_{\downarrow}(r)\\
 v_{\downarrow}(r)\\
 v_{\uparrow}(r)\\
\end{array}\right).
\end{eqnarray}
With the above trial wavefunction, the eigen equation is $H_s(r,\theta,\partial_r,\partial_\theta)\Psi(r,\theta)=E\Psi(r,\theta)$ and the matrix form reads,
\begin{eqnarray}
\left(\begin{array}{cccc}
-\mu+B & P_{-} & \Delta(r,\theta) & 0\\
P_{+} & -\mu-B & 0  & \Delta(r,\theta) \\
\Delta^\dag(r,\theta) & 0 & \mu+B & -P_{-} \\
0 & \Delta^\dag(r,\theta)& -P_{+} & \mu-B  \\
\end{array}\right)\frac{e^{i\nu\theta}}{\sqrt{r}}\left(\begin{array}{c}
e^{i\frac{\pi}{4}}u_{\uparrow}(r)\\
e^{i\theta-i\frac{\pi}{4}}u_{\downarrow}(r)\\
e^{-i\theta+i\frac{\pi}{4}}v_{\downarrow}(r)\\
e^{-i\frac{\pi}{4}}v_{\uparrow}(r)\\
\end{array}\right)&=&E\frac{e^{i\nu\theta}}{\sqrt{r}}\left(\begin{array}{c}
e^{i\frac{\pi}{4}}u_{\uparrow}(r)\\
e^{i\theta-i\frac{\pi}{4}}u_{\downarrow}(r)\\
e^{-i\theta+i\frac{\pi}{4}}v_{\downarrow}(r)\\
e^{-i\frac{\pi}{4}}v_{\uparrow}(r)\\
\end{array}\right),
\end{eqnarray}
where $P_{-}=-ie^{-i\theta}\partial_r-\frac{e^{-i\theta}}{r}\partial_{\theta}$ and $P_{+}= -ie^{i\theta}\partial_r+\frac{e^{i\theta}}{r}\partial_{\theta}$. From the above eigenvalue equation, we can get,
\begin{eqnarray}
(-\mu+B)\frac{u_{\uparrow}}{\sqrt{r}}+v_F(-\partial_r-\frac{\nu+1}{r})\frac{u_{\downarrow}}{\sqrt{r}}+\Delta_0f(r)\frac{v_{\downarrow}}{\sqrt{r}}&=&E\frac{u_{\uparrow}}{\sqrt{r}},\\
v_F(\partial_r-\frac{\nu}{r})\frac{u_{\uparrow}}{\sqrt{r}}-(\mu+B)\frac{u_{\downarrow}}{\sqrt{r}}+\Delta_0f(r)\frac{v_{\uparrow}}{\sqrt{r}}&=&E\frac{u_{\downarrow}}{\sqrt{r}},\\
\Delta_0f(r)\frac{u_{\uparrow}}{\sqrt{r}}+(\mu+B) \frac{v_{\downarrow}}{\sqrt{r}}+v_F(\partial_r+\frac{\nu}{r})\frac{v_{\uparrow}}{\sqrt{r}}&=&E\frac{v_{\downarrow}}{\sqrt{r}},\\
\Delta_0f(r)\frac{u_{\downarrow}}{\sqrt{r}}-v_F(\partial_r-\frac{\nu-1}{r})\frac{v_{\downarrow}}{\sqrt{r}}+(\mu-B)\frac{v_{\uparrow}}{\sqrt{r}}&=&E\frac{v_{\uparrow}}{\sqrt{r}}.
\end{eqnarray}
Now the radial equations can be further written as,
\begin{eqnarray}
\left(\begin{array}{cccc}
-\mu+B & v_F(-\partial_r-\frac{\nu+1}{r}) & \Delta_0f(r) & 0\\
v_F(\partial_r-\frac{\nu}{r}) & -\mu-B & 0  & \Delta_0f(r)  \\
\Delta_0f(r) & 0 & \mu+B & v_F(\partial_r+\frac{\nu}{r}) \\
0 & \Delta_0f(r) & -v_F(\partial_r-\frac{\nu-1}{r}) & \mu-B  \\
\end{array}\right)\frac{1}{\sqrt{r}}\left(\begin{array}{c}
u_{\uparrow}(r)\\
u_{\downarrow}(r)\\
v_{\downarrow}(r)\\
v_{\uparrow}(r)\\
\end{array}\right)&=&E\frac{1}{\sqrt{r}}\left(\begin{array}{c}
 u_{\uparrow}(r)\\
 u_{\downarrow}(r)\\
 v_{\downarrow}(r)\\
 v_{\uparrow}(r)\\
\end{array}\right).
\end{eqnarray}
Here we notice that Hamiltonian matrix is not symmetric. For a Majorana state, its antiparticle is itself and the corresponding wavefunction should satisfy $\mathcal{C}\Psi \propto\Psi$, which leads to $\nu=0$.

 We define $\rho=\frac{r}{\xi_0}=\frac{r}{\hbar v_F/\Delta_0}$ and follow the above definition by setting $\hbar=1$ and we further have $\frac{d}{dr}=\frac{\Delta_0}{v_F}\frac{d}{d\rho}$ and $\frac{d}{dr}\frac{h}{\sqrt{r}}=\frac{\partial_r h}{\sqrt{r}}-\frac{1}{2r}\frac{h}{\sqrt{r}}$. Therefore, the eigenfunction can be further written as,
\begin{eqnarray}
\left(\begin{array}{cccc}
-\bar{\mu}+\bar{B} & -\partial_\rho-\frac{\nu+\frac{1}{2}}{\rho} &  f(\rho) & 0\\
\partial_\rho-\frac{\nu+\frac{1}{2}}{\rho} & -\bar{\mu}-\bar{B} & 0  &  f(\rho)  \\
 f(\rho) & 0 & \bar{\mu}+\bar{B} & \partial_\rho+\frac{\nu-\frac{1}{2}}{\rho} \\
0 &  f(\rho) & -\partial_\rho+\frac{\nu-\frac{1}{2}}{\rho} & \bar{\mu}-\bar{B}  \\
\end{array}\right)\left(\begin{array}{c}
u_{\uparrow}(\rho)\\
u_{\downarrow}(\rho)\\
v_{\downarrow}(\rho)\\
v_{\uparrow}(\rho)\\
\end{array}\right)&=&\bar{E}\left(\begin{array}{c}
 u_{\uparrow}(\rho)\\
 u_{\downarrow}(\rho)\\
 v_{\downarrow}(\rho)\\
 v_{\uparrow}(\rho)\\
\end{array}\right),
\end{eqnarray}
with $\bar{\mu}=\mu/\Delta_0$, $\bar{B}=B/\Delta_0$ and $\bar{E}=E/\Delta_0$.

When discretizing a Dirac equation on a lattice one encounters the problem of fermion doubling. One standard approach is to use a forward-backward difference scheme for approximating the partial derivatives in the above equations\cite{Susskind1977,Stacey1982,Gutierrez2018},
\begin{eqnarray}
\partial_\rho u_{\downarrow}\approx\frac{u_{\downarrow}(\rho+h)-u_{\downarrow}(\rho)}{h} \quad (u_{\downarrow}\rightarrow v_{\downarrow}),\\
\partial_\rho u_{\uparrow}\approx\frac{u_{\uparrow}(\rho)-u_{\uparrow}(\rho-h)}{h} \quad (u_{\uparrow}\rightarrow v_{\uparrow}),
\end{eqnarray}
with $h$ being the discretization step. Here we use the same differential form for $u_{\sigma}$ and $v_{\sigma}$ to preserve the particle-hole symmetry. With discretization on 1D radial geometry with radius $R$, the above equation can be written as,
\begin{eqnarray}
\left(\begin{array}{cccccccc}
-\bar{\mu}+\bar{B} \!& \frac{1}{h}-\frac{\nu+\frac{1}{2}}{\rho_j} &  f(\rho_j) & 0 & 0 & -\frac{1}{h} & 0 & 0 \cdots\\
\frac{1}{h}-\frac{\nu+\frac{1}{2}}{\rho_j} & -\bar{\mu}-\bar{B} & 0  &  f(\rho_j) & 0 & 0 & 0 & 0\cdots  \\
 f(\rho_j) & 0 & \bar{\mu}+\bar{B} & \frac{1}{h}+\frac{\nu-\frac{1}{2}}{\rho_j}& 0 & 0 & 0 & 0\cdots \\
 0 &  f(\rho_j) & \frac{1}{h}+\frac{\nu-\frac{1}{2}}{\rho_{j}} & \bar{\mu}-\bar{B}  & 0 & 0 & -\frac{1}{h} & 0\cdots\\
 0 & 0 & 0 & 0 &-\bar{\mu}+\bar{B} & \frac{1}{h}-\frac{\nu+\frac{1}{2}}{\rho_{j+1}} &  f(\rho_{j+1}) & 0\cdots \\
 -\frac{1}{h} & 0 & 0 & 0 & \frac{1}{h}-\frac{\nu+\frac{1}{2}}{\rho_{j+1}} & -\bar{\mu}-\bar{B} & 0  &  f(\rho_{j+1})\cdots   \\
 0 & 0 & 0 & -\frac{1}{h}& f(\rho_{j+1}) & 0 & \bar{\mu}+\bar{B} & \frac{1}{h}+\frac{\nu-\frac{1}{2}}{\rho_{j+1}}\cdots\\
 0 & 0 & 0  & 0 & 0 &  f(\rho_{j+1}) & \frac{1}{h}+\frac{\nu-\frac{1}{2}}{\rho_{j+1}} & \bar{\mu}-\bar{B}\cdots  \\
 \vdots & \vdots & \vdots & \vdots & \vdots & \vdots & \vdots & \vdots \\
\end{array}\right)\left(\begin{array}{c}
u_{\uparrow}(\rho_j)\\
u_{\downarrow}(\rho_j)\\
v_{\downarrow}(\rho_j)\\
v_{\uparrow}(\rho_j)\\
u_{\uparrow}(\rho_{j+1})\\
u_{\downarrow}(\rho_{j+1})\\
v_{\downarrow}(\rho_{j+1})\\
v_{\uparrow}(\rho_{j+1})\\
\vdots\\
\end{array}\right)\!\!\!\!&=&\!\!\!\!\bar{E}\left(\begin{array}{c}
u_{\uparrow}(\rho_j)\\
u_{\downarrow}(\rho_j)\\
v_{\downarrow}(\rho_j)\\
v_{\uparrow}(\rho_j)\\
u_{\uparrow}(\rho_{j+1})\\
u_{\downarrow}(\rho_{j+1})\\
v_{\downarrow}(\rho_{j+1})\\
v_{\uparrow}(\rho_{j+1})\\
\vdots\\
\end{array}\right).
\end{eqnarray}
The above matrix has a general form as,
\begin{eqnarray}
&&\left(\begin{array}{ccccc}
H_{00}(\rho_j) & H_{01} & 0& 0&  \cdots\\
H^\dag_{01} & H_{00}(\rho_{j+1}) &  H_{01}  & 0&  \cdots  \\
0 & H^\dag_{01} & H_{00}(\rho_{j+2}) &  H_{01}  &   \cdots  \\
 \vdots & \vdots & \vdots & \vdots& \vdots\\
\end{array}\right)\left(\begin{array}{c}
\psi(\rho_j)\\
\psi(\rho_{j+1})\\
\psi(\rho_{j+2})\\
\vdots\\
\end{array}\right)=\bar{E}\left(\begin{array}{c}
\psi(\rho_j)\\
\psi(\rho_{j+1})\\
\psi(\rho_{j+2})\\
\vdots\\
\end{array}\right)\\
H_{00}(\rho_j)&=&\left(\begin{array}{cccc}
-\bar{\mu}+\bar{B} & \frac{1}{h}-\frac{\nu+\frac{1}{2}}{\rho_j} &  f(\rho_j) & 0  \\
\frac{1}{h}-\frac{\nu+\frac{1}{2}}{\rho_j} & -\bar{\mu}-\bar{B} & 0  &  f(\rho_j)    \\
 f(\rho_j) & 0 & \bar{\mu}+\bar{B} & \frac{1}{h}+\frac{\nu-\frac{1}{2}}{\rho_j}  \\
 0 &  f(\rho_j) & \frac{1}{h}+\frac{\nu-\frac{1}{2}}{\rho_{j}} & \bar{\mu}-\bar{B} \\
\end{array}\right)\\
H_{01}&=&\left(\begin{array}{cccc}
 0 & -\frac{1}{h} & 0 & 0  \\
 0 & 0 & 0 & 0  \\
 0 & 0 & 0 & 0 \\
 0 & 0 & -\frac{1}{h} & 0 \\
\end{array}\right),
\end{eqnarray}
where $h=\frac{R-R_{min}}{N-1}$ and $\rho_j=R_{min}+(j-1)h$ with $j=1,2,...,N$. In the calculations, we adopt $R_{min}=0.01$, $N=1001$ and $R=50$. For the calculations of CMMs, the Zeeman field is assumed to be,
\begin{equation}
B(r)=
\begin{cases}
B_0& r\leq R_0,\\
0& r>R_0,
\end{cases}
\end{equation}
with $B_0=2.2>\sqrt{\mu^2+\Delta^2_0}$.  After solving the eigenvalue equation, we calculate the local density of states (LDOS) to simulate the tunneling conductance measured by STM using,
\begin{eqnarray}
\frac{dI}{dV}(r,E) \propto LDOS(r,E)=\frac{1}{r}\sum_{\nu n\sigma}[|u^{\nu}_{n\sigma}(r)|^2\delta(E-E^{\nu}_n)+|v^{\nu}_{n\sigma}(r)|^2\delta(E+E^{\nu}_n)].
\end{eqnarray}
The spectrum of a vortex in the superconducting Dirac state is displayed in Fig.\ref{mzm_radial}(a). We discard the artificial CMM localized at the outer boundary of the disk. The pink circles denote the bound states inside the vortex and the zero-energy state is the Majorana mode. With including a Zeeman filed $B$ for $|r|\le R_0=5$ region, the spectrum is displayed in Fig.\ref{mzm_radial}(b) and the local Majorana mode of the vortex transforms into a CMM (orange circles) localized at the domain wall. The energy quantum of the CMM is proportional to $\frac{1}{L}$ with $L$ being the circumference of the region with exchange field, as shown in Fig.\ref{mzm_radial}(c). 

\begin{figure}[tb]
\centerline{\includegraphics[width=0.8\columnwidth]{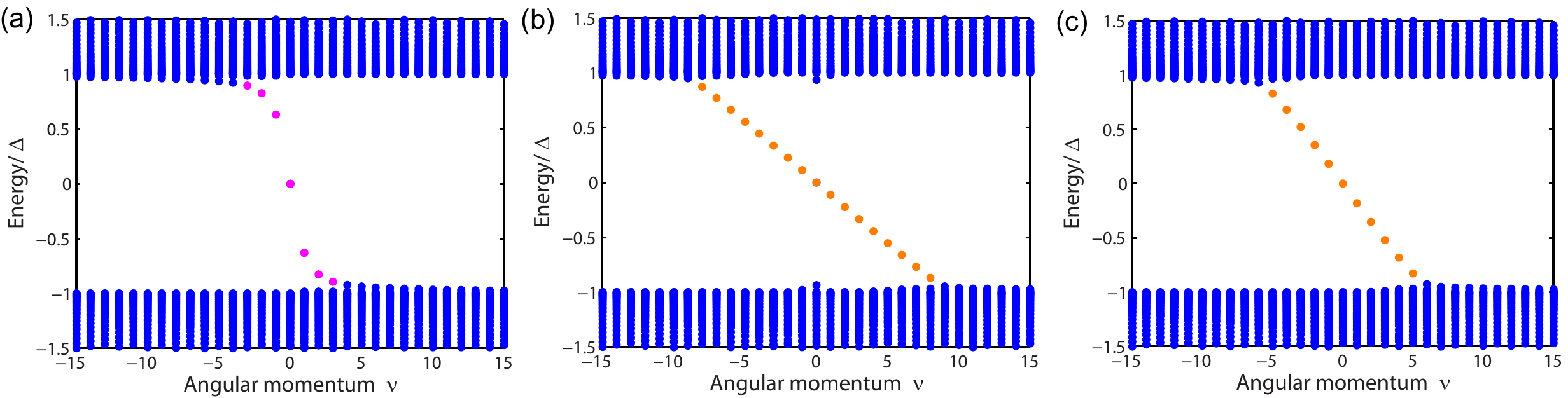}}
\caption{(color online) Energy spectra for a vortex in the superconducting Dirac surface states: (a) $B=0$, $B(r)=B_0\Theta(R_0-r)$ (b) $R_0=5$ and (c) $R_0=3$. The adopted chemical potential $\mu=1$ and $\Delta(r)=\Delta_0 \tanh\frac{r}{\xi_0}$ with $\Delta_0=1$ and $\xi_0=1$. The profiles of MZMs as a function of Zeeman field along $z$ direction.
 \label{mzm_radial} }
\end{figure}

  \end{widetext}

\end{document}